\documentclass[preprint,prl,aps,floats]{revtex4}
\usepackage{graphicx}
\usepackage{amsmath}
\begin{document}

\title{Solution of the Dirac Equation using the Lanczos Algorithm}
\author{R. C. Andrew, H. G. Miller\footnote{E-Mail:hmiller@maple.up.ac.za} }
\affiliation{Department of Physics, University of Pretoria, Pretoria 0002, South Africa}
\author{and G. D. Yen}
\affiliation{Department of Information Management, WuFeng Institute of Technology, Minsyung, Jiayee 621-53, Taiwan}
\begin{abstract}
\noindent

Covergent eigensolutions of the Dirac Equation for a relativistic electron in an external Coulomb potential are obtained using the Lanczos Algorithm. A tri-diagonal matrix representation of the Dirac Hamiltonian operator is constructed iteratively and diagonalized after each iteration step to form a sequence of convergent eigenvalue solutions.  Any spurious solutions which arise  from the presence of continuum states can easily be identified.

\noindent
PACS\ 03.65.Ge,\ 02.60.Lj 

\end{abstract}
\maketitle
\section{Introduction}
In relativistic quantum mechanics, the calculation of the lowest lying bound states of a
relativistic electron in a static Coulomb potential is a classic problem. In some respects the perturbative solution provided by Bethe\cite{B64} is both simple and elegant and a standard example in many textbooks.  Early attempts to obtain the exact eigensolutions\cite{D75,GMP76} employed finite difference methods to solve the radial equations and obtain the eigenenergies.   Recently much effort has focused on the extension of basis set methods\cite{KH02,DG81,GD82,GD88} which have been applied with success in  non-relativistic quantum mechanics. Such matrix approximations of the Dirac equation are  usually  obtained from a variational principle with the radial components approximated by a finite number of terms of a basis-set expansion. They often display pathological features such as the occurrence of unphysical spurious solutions\cite{K84,KH02} and a failure to vary systematically with the basis-set size. These can for the most part be avoided if a resolvent operator rather than the Dirac Hamiltonian is used in the variational principle\cite{KH02} or by construction in explicit finite basis-set calculations\cite{G85}. Accurate numerical solutions have also recently been obtained using mapped Fourier grid methods\cite{AH05,BRB03}. Spurious roots are still present in certain cases but can easily be identified\cite{AH05}.

For the Coulomb potential with a coupling constant which is less than $\sqrt{\frac{3}{2}}$, the Dirac Hamiltonian is self-adjoint in the  first Sobolev space\cite{T92}. Hence, as we shall show, its eigenpairs can be obtained iteratively  in an absolutely convergent manner
by means of the Lanczos Algorithm\cite{L50,KMB81} provided, of course, that only the positive energy eigensolutions are considered. Although the Lanczos Algorithm is generally known for its application to conventional matrix eigenvalue problems, it can be used to solve eigenvalue problems of a self-adjoint operator which is bounded from below or above\cite{KMB81}.  In the former case, the matrix to be diagonalized is iteratively transformed into a tri-diagonal matrix and diagonalized after each iteration step to form a sequence of convergent eigenvalue solutions.  In the latter case, a tri-diagonal representation of the self-adjoint operator is constructed iteratively and diagonalized after each iteration step to form a sequence of convergent eigenvalue solutions.  Any spurious solutions which arise  from the presence of continuum states can easily be identified in manner recently suggested\cite{AM03}. Furthermore, many  of the aforementioned problems are avoided if the following iterative scheme proposed by Lanczos is employed.

\section{The Lanczos Method}
Given the following eigenvalue problem
\begin{equation}
 \hat{H}|E_{\lambda}\rangle=E_{\lambda}|E_{\lambda}\rangle
\end{equation}
where $  \hat{H}  $ is a self-adjoint operator (not necessarily bounded) in a
separable Hilbert (or Sobolev) space which possesses a number
of eigenvalues $  E_{\lambda }  $ (not necessarily bounded from above) which ascend
from a minimal one $  E_{1 }  $.  Furthermore let $  |1\rangle  $ be a
normalized start vector having the properties that:
\begin{enumerate}
\item $  \hat{H}^{k }|1\rangle  $  exists for all non-negative integers k and
\item $  \langle x|1\rangle  $ satisfies the boundary conditions.
\end{enumerate}
The algorithm may then be stated as follows. Calculate successively the
vectors
\begin{equation}
|\phi_{1 }\rangle:=|1\rangle
\end{equation}
\begin{equation}
|\phi_{n +  1 }\rangle:=\frac{1}{||\ldots||}(\hat{H}^{n }|1
\rangle-\sum\limits_{n'=1 }^{n }|\phi_{n' }\rangle
\langle \phi_{n' }|\hat{H}^{n } |1\rangle), \hspace{5mm}
n=1,2,3,\ldots
\end{equation}
where $n$ is the iteration number. The Lanczos Method solves the eigenproblem by diagonalizing successively the operators
\begin{equation}\label{diag} 
\hat{H}_{n }:=\sum\limits_{m,m'=1 }^{n }
|\phi_{m }\rangle\langle\phi_{m }|\hat{H}|\phi_{m' }
\rangle\langle\phi_{m' }|, \hspace{10mm} n=1,2,3,\ldots
\end{equation}
The orthonormalized Lanczos vectors $  |\phi_{n }\rangle  $ fulfill the following recursion relations~\cite{KMB81}
\begin{equation}
|\phi_{n }\rangle=p_{n}(\hat{H})|1\rangle
\end{equation}
where the Lanczos polynomials $  p_{n }(x)  $ are defined in the
following way:
\begin{equation}
p_{1 }(x):=1 \hspace{15mm} p_{2 }(x):= w_{1}^{-1 }(x-v_{1 }),
\end{equation}
\begin{equation}
p_{n+1 }(x):=w_{n }^{-1 } [ (x-v_{n } )p_{n }(x)
-w_{n-1 }p_{n-1 }(x)] \hspace{5mm} n=2,3,4,\ldots,
\end{equation}
with
\begin{equation}
v_{n }:= \langle\phi_{n }|\hat{H}|\phi_{n }\rangle
\end{equation}
\begin{equation}
w_{n}:= \langle\phi_{n+1 }|\hat{H}|\phi_{n }\rangle.
\end{equation}

Performing the diagonalizations in equation \ref{diag} reduces to finding the roots of the following characteristic polynomial
\begin{equation}
det(\hat{H}_{n }-x\cdot\hat{1}):=(-1)^{n }w_{1 }\ldots w_{n }p_{n+1 }(x)
\end{equation}
after each iteration step.

The generated sequence of eigenpairs ($e_{\lambda\,n }$ , $|e_{\lambda\,n }\rangle$) possess the following convergence
properties~\cite{KMB81}
\begin{equation}
e_{\lambda\,n }
\begin{array}{cc}
\longrightarrow &    \\
n\to\infty  &
\end{array} E_{\lambda }, \hspace{10mm} \lambda=1,2,3,\ldots,
\end{equation}
\begin{equation}
|e_{\lambda\,n } \rangle
\begin{array}{cc}
\longrightarrow &    \\
n\to\infty  &
\end{array} |E_{\lambda }\rangle, \hspace{10mm} \lambda=1,2,3,\ldots .
\end{equation}

 For a self-adjoint operator the Lanczos algorithm can be implemented using \textit{Mathematica}\cite{W91} 

\section{Identifying spurious solutions}
 The exact bound states can be identified in the following manner\cite{AM03}.
For those  operators which possess  a continuum as well as a point 
spectrum, the space spanned  by the bound state eigenfunctions is by itself 
certainly not complete and  a suitable  start vector should  be composed 
only of components in the subspace spanned by the bound state eigenvectors. 
  Usually the start vector is chosen from a 
complete set of analytic functions which define a space 
$\mathcal{F}$. This space is in most cases not necessarily of the same dimension 
as the subspace spanned by the exact eigenvectors. On the other hand, if the 
Lanczos algorithm is applied with this  choice  for the start vector, the 
eigenpairs obtained will correspond to those of the operator  projected 
onto $\mathcal{F}$. A subset of these eigenstates must correspond to the exact 
eigenpairs of the unprojected Hamiltonian operator since the exact eigenstates 
can be expanded in terms of the complete set of states  which span  
$\mathcal{F}$.   After  
each iteration,  for each of the converging eigenpairs ($e_{\lambda\,n}$, $|e_{\lambda\,n}\rangle$),
\begin{equation}
\Delta_{\lambda\,n}=|e_{\lambda\,n}^2-\langle e_{\lambda\,n}|\hat{H}^2|e_{\lambda\,n}\rangle| \label{delta}
\end{equation} 
(where $n$ is the iteration number) is 
calculated and a determination is made  as to whether  $\Delta$ is converging  
toward zero or not.  For the exact bound states of $\hat{H}$,  $\Delta$ must be 
identically zero while the  other eigenstates states of the projected operator 
should converge  to some non-zero positive   value. Provided sufficient 
iterations are performed, it should be possible, in this manner to identify 
uniquely the approximate eigenpairs which ultimately will converge to the exact 
bound states. Since in the the Lanczos algorithm the lowest-lying eigenpairs 
(for an operator which possesses at most a lower bound) usually converge the 
quickest\cite{P80} one should be able to obtain good approximations to the 
lowest-lying eigenpairs.

\section{The Dirac Equations}
The coupled radial Dirac equations for the upper and lower components G(r) and F(r) can be written as\cite{S84}
\begin{equation}
-\hbar c \frac{dF(r)}{dr} - \frac{(1-\kappa)\hbar c}{r} F(r)= (E-V(r) - m c^2)G(r), 
\end{equation}
\begin{equation}
\hbar c \frac{dG(r)}{dr} + \frac{(1+\kappa)\hbar c}{r} G(r)= (E-V(r) + m c^2)F(r),
\end{equation}
where $\kappa$ is a quantum number analogous to the angular momentum quantum number $\mathit{l}$ in non-relativistic quantum mechanics. Using $\mathit{f(r)=rF(r),\ g(r)=rG(r)}$ and the Coulomb potential for charge Z, in relativistic units $\mathit{\hbar=m_e=c=1}$ the equations become 
\begin{equation}
-\frac{df(r)}{dr}+\frac{\kappa}{r}f(r)=(E+\frac{z\alpha}{r}-1)g(r), \label{f}
\end{equation}
\begin{equation}
\frac{dg(r)}{dr}+\frac{\kappa}{r}g(r)=(E+\frac{z\alpha}{r}+1)f(r), \label{g}
\end{equation}
where $\mathit{\alpha}$ is the fine structure constant. Equations (\ref{f}) and (\ref{g}) can written in matrix notation
\begin{equation}
\left(
\begin{tabular}{cc}
$-(\frac{z\alpha}{r}-1)$ & $-(\frac{d}{dr}-\frac{\kappa}{r})$\cr
$(\frac{d}{dr}+\frac{\kappa}{r})$ &$ -(\frac{z\alpha}{r}+1) $\cr
\end{tabular} 
\right)
\left(
\begin{tabular}{c}
$g(r)$\cr
$f(r)$\cr
\end{tabular}
\right)=E
\left(
\begin{tabular}{c}
$g(r)$\cr
$f(r)$\cr
\end{tabular}
\right)
\end{equation}
\begin{equation}
\hat{H}_{D}
|\Psi\rangle=
E|\Psi\rangle \label{dham}
\end{equation}
where 
\begin{equation}
|\Psi\rangle=
\left(
\begin{tabular}{c}
$g(r)$\cr
$f(r)$\cr 
\end{tabular}
\right)
\end{equation}
 Equation (\ref{dham}) defines the eigenvalue problem which will be solved iteratively using the Lanczos algorithm.

\section{Numerical results}

To demonstrate that the the Lanczos algorithm may be used to  solve the  Dirac equations, a start vector was chosen using a slightly perturbed ground state solution derived by Bethe\cite{B64} with ($\kappa\,=\,-1$) for an atom with $z\,=\,100$.   
The start vector contained the following perturbed radial functions

\begin{equation}
g_{1}(r)=(r^{s})^{\gamma}e^{-z\alpha r}
\end{equation}
\begin{equation}
f_{1}(r)=\frac{(s-1)(r^{s})^{\gamma}e^{-z\alpha r}}{z\alpha}
\end{equation}
where $s=\sqrt{1-(z\alpha)^2}$ and a perturbation factor $\gamma=.0998$. Note this choice for the start vector eliminates the possibility of obtaining negative energy eigensolutions\cite{B64}.
\begin{figure}
\centering
\includegraphics[angle=-90,scale=0.5]{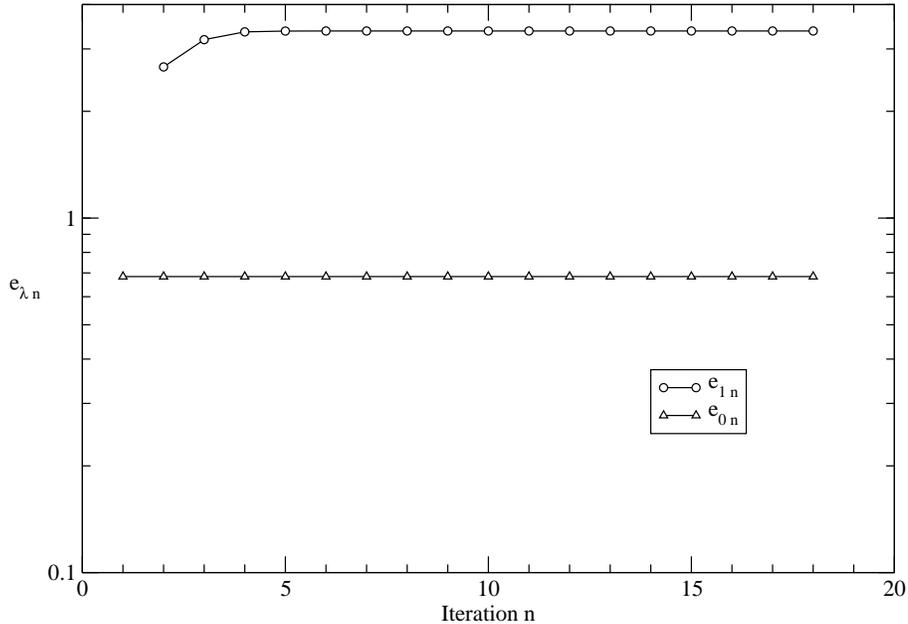}
\label{fig1}
\caption{ $e_{\lambda\,n}$ vs. n for the bound state and the spurious solution}
\end{figure}
\begin{figure}
\centering
\includegraphics[angle=-90,scale=0.5]{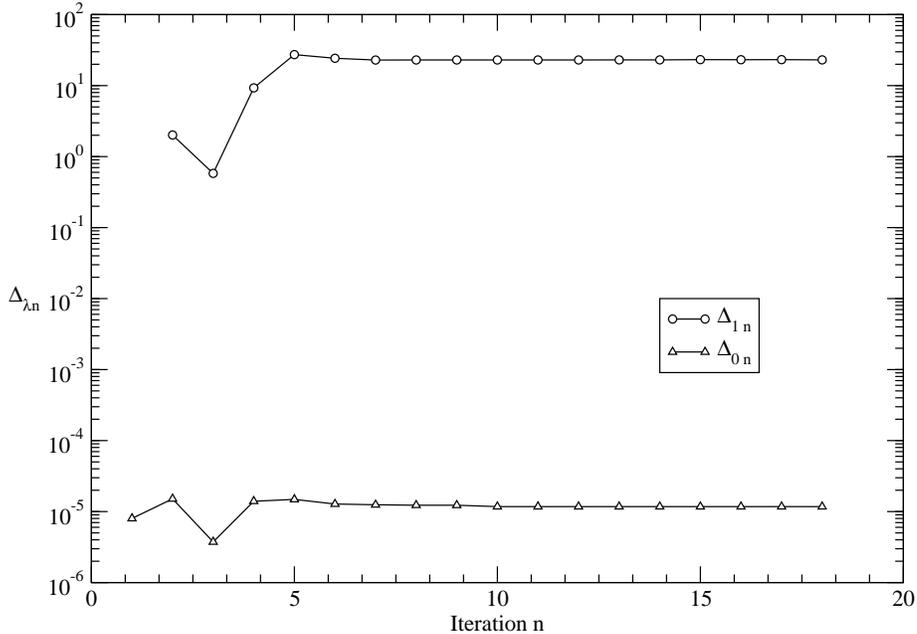}
\label{fig2}
\caption{ $\Delta_{\lambda, n}$ vs. n for the bound state and the spurious solution}
\end{figure}

Figure 1 shows the convergence for the ground state eigenvalue $e_{o\,n}$ and a spurious solution $e_{1\,n}$. After 18 iterations, $e_{o\,n}$ quickly converged to the exact ground state value of $E_{o}=\simeq0.683729$. The unphysical spurious solution converged to a value of 3.37003. Using a larger perturbation factor of $\gamma=.95$, $e_{0\,n}$ converged to $E_{0}$ to within two significant figures after 18 iterations. 

Figure 2 shows the convergence of $\Delta_{\lambda,n}$ for the ground state solution and the spurious solution. $\Delta_{\lambda,n}$ iteratively converges to zero for the generated bound state solution and converges to a non-zero value for the spurious solution.

\section{Conclusions}
The eigensolutions of the Dirac Hamiltonian have successfully been obtained using the  the Lanczos algorithm.  An eigensolution can be identified by testing the convergence of $\Delta_{\lambda,n}$ for each generated eigenpair and by discarding generated solutions for which $\Delta$ does not converge to zero. The convergence rate, needless to say , depends on the choice of the initial start vector.  

Furthermore, it has been pointed out \cite{P80} that the Lanczos algorithm may 
be regarded as an application of the Rayleigh-Ritz method in which an 
orthonormalized set of Krylov vectors
have been used to define the successive subspaces in which the operator is diagonalized. 
This being the case, the convergence of $\Delta$ 
should also be a suitable means of sorting out the approximate exact eigenstates
from the 
spurious eigenstates if the Rayleigh-Ritz method is employed.


\end{document}